\documentclass[12pt]{article}
\usepackage{graphicx,amsmath}
\title{{What's wrong with the Platonic ideal of space and time?}}

\author{Lorenzo Sadun\thanks{Department of Mathematics, University of Texas, Austin, TX 78712; sadun@math.utexas.edu} 
}

\begin{document}
\maketitle

To our senses, space is smooth, 3-dimensional, and flat. We move in a
continuum where all points are equal (space is ``homogeneous'') and
all directions are equal (``isotropic'', or ``round''). If we head off
in any direction, we keep on going, with no curving back on ourselves
(``flat''). In short, we seem to live in a universe governed by
Euclidean geometry.

Actually, almost everything I wrote in the previous paragraph was a
lie. Our senses continually detect the difference between different
directions. Things fall down, not up. The sun rises in the East, not
in the West. We {\em know} that not all points are equal, and that
Hawaii is a lot more pleasant than Antarctica.  Curvature is all
around us, from the hill my house sits on to the fact that we can fly
around the world.

All the same, most of us still believe in 3-dimensional space,
plus an added dimension (time) that describes how things change. We
rely on a mathematical model of reality in which space is a Platonic
ideal: smooth, homogeneous, isotropic, and flat. Everything that
breaks that underlying symmetry is attributed to objects: a planet
whose gravity causes things to fall in a preferred direction, and
whose rotation makes another object appear to move through
the sky, the hills and valleys of my home town,
and lovely tropical islands.  We think that space is simple, that objects are
complicated, and that the job of scientists is to understand the messy
behavior of objects against the perfect backdrop of space.

This idealization of perfect space and imperfect contents 
makes for a lovely theory,
but is it correct? It was accepted almost without question for over 20
centuries, from the ancient Greeks through the Middle Ages and the
Enlightenment. In the late 19th century, however, it started to break
down. While the theory works very well to describe physics on many
length scales, it gives results that are nonsensical, or at least that
contradict experiment, when dealing with very big things, very small
things, and very fast things.

In this essay, I'll touch briefly on the new theories that were
developed to deal with these discrepancies --- special relativity,
general relativity, quantum mechanics and string theory. I'll then
turn to the suggestion, popularized by Stephen Wolfram \cite{Wolfram}, 
that space
and time aren't smooth at all, but come in essentially discrete
chunks. Using recent results from the theory of aperiodic tilings,
I'll argue that this last suggestion is not realistic, and will defend
the conventional wisdom that Euclidean space and time, as modified
slightly by 20th century physics, is still the best way to describe
reality.

\section{Relativity and the fall of Euclidean space}

If space and time are absolutes, then how fast are we moving? After
all, the earth rotates on its axis and revolves around the sun, the sun
revolves around the center of the galaxy, and the galaxy tumbles
through the universe. We {\em must} be moving, but in what direction,
and how fast?  In 1887, Michelson and Morley \cite{MM} tried to find out. They
figured that light moving in the same direction as us would appear to
be moving slower (since it has to catch up with us), that light moving
in the opposite direction would appear to be moving faster, and that
light moving perpendicular to our motion would have an intermediate
speed. With a clever interferometry experiment, they measured these
differences and got exactly zero, suggesting that we were not moving
at all! How could that be?

Several complicated mechanisms were proposed for why the speed of
light {\em appeared} to be the same in all directions. It took almost
20 years, until Einstein's 1905 Special Theory of Relativity\cite{special}, for
mankind to realize that the laws of mechanics and electromagnetism,
and hence the speed of light, really were the same relative to the
earth, to the sun, and to the distant galaxies. Different observers
moving relative to one another have different notions of space and
different notions of time, but the same laws of physics. Space and
time are not absolutes, but are only defined relative to an observer.
Moreover, they are closely linked, and switching from one reference
frame to another is like a rotation in a 4-dimensional
space-time\footnote{When you rotate in the $x$-$y$ plane, the new value of
  $x$ depends on both the old values of $x$ and the old value of
  $y$. Likewise, when you do a ``Lorentz transformation'' from one
  reference frame to another, the new position depends on both the old
  position and the old time, as does the new time.} As such, it is
impossible to speak of the nature of space without also considering
the nature of time, and vice-versa. To know one is to know both.
 
Einstein took things a step further in his 1915 General Theory of
Relativity (GR) \cite{GR}.  He proposed that space-time is not flat. Rather, the
presence of mass, momentum and energy causes space to bend, and we
perceive this bending as gravity.  We can no longer place perfect
space and imperfect matter in separate categories. Matter bends space
and the geometry of space affects matter. If the distribution of
matter isn't uniform and isotropic, then neither is the geometry of
space-time. {\em Matter is lumpy, so space-time is bumpy.}
  
Since that time, GR has been tested in numerous
experiments, and has performed extremely well, most recently in the 2016 
observation of gravitational radiation. GR may not be the
ultimate theory and may require tweaking in the future (In particular,
Andrei Sakharov \cite{Sakharov} has argued that it is just the first term in an infinite
series of corrections to Newton's Laws), but it is hard to avoid the
conclusion that, on extremely large length scales, 
the Platonic ideal of space-time just doesn't work.
 
\section{Quantum mechanics and the very small}
  
A different challenge to classical physics came when studying very
small distances. According to classical physics, a glowing hot object
should emit a certain amount of long-wavelength infrared light. It
should emit more shorter-wavelength visible light, still more
ultraviolet light, yet more x-rays, and so on. Not only is the bulk of
the radiation supposed to be of such high frequency that a coal from
your backyard grill would kill you, but the total amount of energy
emitted per unit time is supposed to be {\em infinite}.

To explain why glowing coals aren't lethal, Max Planck \cite{Planck}
proposed that light
energy can only be emitted or absorbed in discrete chunks, called quanta. 
This theory of light, called quantum mechanics, was soon extended to all forms
of matter and energy and then generalized to fields that describe the 
creation and annihilation of particles. This body of work took care of the
``ultraviolet catastrophe'' that puzzled Planck, but created other mysteries.

For one, Werner Heisenberg \cite{uncertain} 
observed that quantities that were once thought to
be precise, like the position and momentum of a particle, are actually
a bit fuzzy. There is uncertainty to position, there is uncertainly to
momentum, and the product of the two uncertainties is at least Planck's
constant divided by $4\pi$. Likewise, there is uncertainly in energy
and uncertainty in the time when things happen. By general relativity, the
curvature of space is a function of mass and energy and momentum, but these
quantities can't be nailed down. So not only are particles fuzzy, but 
space-time itself is fuzzy. 

Worse still, the infinities that appeared in the ultraviolet
catastrophe aren't completely tamed. By the uncertainty principle in
energy, particles can blink in and out of existence. In many problems
in quantum field theory, the effect of all these ``virtual particles''
could be infinite, which doesn't make sense. To avoid these infinites,
the laws of physics, and of geometry, have to become qualitatively
different at the scale of the so-called ``Planck length''. This is an incredibly
small length of around $10^{-35}$ meters, or about a septillionth the
radius of an atomic nucleus. (You could fit more Planck-length sized
particles into a single proton than you could fit protons within a
million-mile diameter ball.)

According to string theory, space-time isn't 4 dimensional. It's
actually 10 dimensional (or 11 dimensional in some versions), with all
but 4 of the dimensions wrapped up in a higher dimensional analogue of
a surface, of size comparable to the Planck
length. Just as we can treat a thin 3-dimensional filament, such as a human
hair, as being
effectively 1-dimensional, our thin 10 or 11 dimensional universe is
effectively 4 dimensional.

A very different solution has been advocated by Stephen Wolfram 
\cite{Wolfram}. He suggests 
that at very small length scales the universe is really 0-dimensional! 
His theory is that space and time are actually discrete, with the possible
points ordered in a neat array. At each new time step, what is 
happening at each point in space depends only on what was happening at that
point, and at all adjacent points, an instant earlier. That is, the universe is
like a gigantic array of computers, each one updating based on what its
neighbors are doing.  

\section{Life on the grid?}

Such an array is called a ``cellular automaton''. The past decades have 
seen an explosion of work on cellular automatons, including notable advances
by Wolfram himself.  The most famous example of a cellular automaton is John
Conway's Game of Life \cite{life}.
This
game operates on a 2-dimensional grid of square ``cells'', and time
advances in discrete steps called ``ticks''. At any given 
time, each cell is either
alive or dead. At each tick of the clock, each live cell either survives or
dies, depending on the number of live cells in the 8 positions around it,
and each dead cell either stays dead or comes to life (``birth'') 
by a slightly different 
rule. This simple game exhibits amazingly complicated behavior, with intricate
patterns propagating across the screen. 

Could a 3-dimensional version of this sort of game be a model for the
complex behavior of the real world?  Wolfram says yes, but I say
no. In the Game of Life, signals propagate at a maximum speed, just
like the speed of light, but this speed depends on direction. Whether
a cell at (0,0) is alive or dead at time 0 affects all the neighboring
cells at time 1, all the cells around those at time 2, and so
on. After $n$ time steps, the cells that are potentially influenced by
the initial situation form a square with vertices at $(n,n)$, $(-n,n)$,
$(-n,-n)$ and $(n,-n)$. Signals propagate fastest in the diagonal
directions and slowest sideways or up-and-down. This contradicts the
experimental fact that the speed of light is the same in all
directions.

You might argue that this contradiction resulted from the details of the
Game of Life, and that different rules might give a different speed of
light. It's true that more complicated rules can make things a {\em bit}\/
more isotropic, but they can't make things completely round. As long as each
cell has a finite number of neighbors, there will always be a finite
number of directions in which information runs fastest, and
intermediate directions in which information runs slower.
Put another way, if the underlying geometry of space-time is a grid, then
there will always be physical phenomena that reveal the underlying axes of 
the grid, in the same way that the facets of a crystal reveal the underlying
arrangement of the atoms inside. 

(An important caveat: Computers use grids to model continuous and
isotropic systems all the time.  However, these numerical models only
work well when looking at patterns that move much slower than the
maximum transmission speed of information, a.k.a.~the speed of
light. Cellular automata can accurately model a world governed by Newton's
laws, and can be very useful in understanding a cold weather front
that is moving at 15 miles per hour, but they can't handle
extreme relativistic motion.)

\section{Life in a raindrop?}

The universe isn't a grid, but can it still be discrete? Just because a
crystal can't be round doesn't mean that we can't make something round 
(or at least round to the naked eye) out of atoms. The raindrops falling
outside my window say that you can!  If you take a bunch of 
building blocks and assemble them randomly, as with the grains of sand in 
a sand pile or the water molecules in a raindrop, 
the resulting structure is unlikely to have any preferred 
directions. 

However, random structures have their own problems. Imagine a small
explosion in the middle of a sand pile. The sound from that explosion
wouldn't go straight to our ears, but instead would richochet off of
the various grains of sound in random directions. The sound {\em
  would} go in all directions at essentially the same speed, but
different paths would take different amounts of time to reach us. What
started out as a sharp BANG!  would be heard as a not-so-sharp
roar. The wave properties of sound (constructive and destructive
interference) could reduce this effect but cannot eliminate it. Waves
of different frequencies work their way through the maze at slightly
different rates; in raindrops, this distortion causes rainbows. Waves
propagating through random media {\em always} get distorted and
smeared.

However, signals from distant galaxies do {\em not} get blurred as they travel
to us through empty space. 
The neutrino bursts from a supernova, or the gravitational waves 
from the merging of two black holes, travel for billions of years across
the universe and then hit us in an instant. We don't see any of the fuzziness
that would be expected from random space-time. 

In addition to the experimental evidence against random discrete
space-time, such a model would raise as many additional metaphysical
questions as it would answer.  What determines the random structure at
each point in space-time? The random arrangements of sand in a sand
pile reflect the details of how the grains of sand were dropped and
mixed, but there is no {\em process} by which space and time are
created. Space and time just {\em are}.  Einstein famously objected to the role
of pure chance in quantum mechanics; this would be far worse.

\section{Life in an aperiodic tiling?}

Finally, we consider a possibility intermediate between random
space-time and a regular periodic grid. It is possible to have order
without periodicity.  For instance, imagine a sequence of $+$ signs and
$-$ signs. We start with a $+$ sign and follow it with its opposite to get
$+-$. We then follow this with the opposite of the pair, namely $-+$,
to get $+--+$. We then follow this with the opposite of $+--+$, namely
$-++-$, to get $+--+-++-$. Continuing the process forever, we get a
infinite sequence, called the Thue-Morse sequence, with the magical property
that no pattern within it (e.g., $+--+$) ever repeats itself 3 times
in a row.  The Thue-Morse sequence is an example of {\em aperiodic
  order}, in which the arrangements follow precise rules but are not
just the same pattern repeated over and over and over.

An interesting 2-dimensional aperiodic tiling is 
the {\em pinwheel tiling} \cite{pinwheel}
invented by John Conway and Charles Radin. The basic tiles are right
triangles with sides of length $1$ and $2$ and hypotenuse of length $\sqrt{5}$.
You can arrange five such tiles to make a bigger triangle of the same
shape, which we call a {\em supertile of level 1}. We can then arrange
five supertiles of level 1 to make a supertile of level 2, 5 of those to 
make a supertile of level 3, and so on. This design is featured architecturally
in Federation Square in Melbourne, Australia, and in the author's home.

\begin{figure}[ht]
\centering
\includegraphics[angle=0,width=0.9\textwidth]{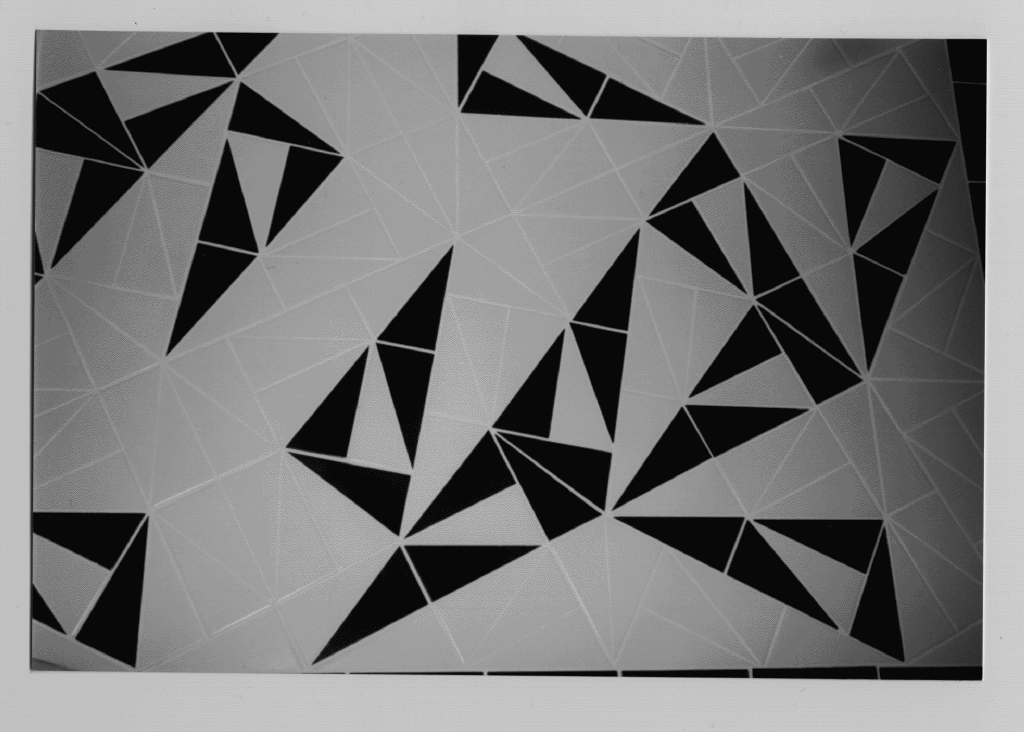} 
\caption{The author's bathroom floor. Note the light level-1 supertile
sitting in the center of a mostly dark level-2 supertile, which is itself
in the center of a level-3 supertile that extends beyond the frame.}
\end{figure}

The center tile of a supertile of level 1 is the same shape as the supertile, 
but is rotated by the angle $\theta_0=\tan^{-1}(1/2)$. If the supertile of 
level 1 is in the center of a supertile of level 2, then the center tile is
rotated by $2 \theta_0$ relative to the level 2 supertile. Continuing the 
process, we get rotations by arbitrary multiples of $\theta_0$.

However, $\theta_0$ is an irrational number of degrees, so no multiple
of $\theta_0$ will ever take you back exactly to the direction you
started in. The pinwheel tiling has tiles pointing in infinitely many
different directions, and all directions are equally likely. (In
technical language, the distribution of directions is {\em uniform} on
the circle.) While the tiles themselves are pointy triangles, the
statistical properties of the pinwheel tiling are rotationally
invariant, with no directions preferred over any others. Could the
pinwheel tiling, or something like it, be a discrete model for a
seemingly isotropic universe?

The problem is that the rotational invariance only manifests itself in
the limit of infinite size, and develops {\em incredibly} slowly. An $n$-th
level supertile has $5^n$ tiles that appear in only $8n$ different
directions, with a still smaller number of directions accounting for 
the vast majority of the tiles.  If the tiles were the size of the
Planck length, then a Milky Way Galaxy-sized supertile might have
$10^{110} \sim 5^{160}$ tiles in it, but the bulk of those tiles would
only be pointing in about 100 different directions. Even at 
astronomical length scales, space would not look isotropic.

Things are qualitatively the same for {\em all} 2-dimensional
hierarchical tilings, and only slightly better in 3 dimensions.  To
get around the 2-dimensional limitations, Conway and Radin devised a
3-dimensional generalization of the pinwheel tiling, called the {\em
  quaquaversal} tiling (Latin for ``every which way'')
\cite{quaqua}. The number of relevant directions does grow faster than
for the pinwheel, but a galaxy-sized supertile would still only
feature a few thousand relevant directions \cite{draco}.

\section{Conclusions}

The Platonic ideal of perfectly uniform and symmetric 3-dimensional space,
coupled with perfectly uniform 1-dimensional time, did not stand up to
20th century physics. Special relativity shows that we can't study space 
and time separately, but must instead think about 4-dimensional space-time. 
General relativity shows that space is not flat, but bends and curves in 
response to the matter that is in it. Quantum mechanics says that this matter
is fundamentally uncertain, making the structure of space-time uncertain. 
Furthermore, something fundamentally different has to happen
at the ultra-microscopic Planck length. 

String theory says that, at the Planck scale, space-time is actually
10 or 11 dimensional, with all but 4 dimensions curled up into a tight
ball. Many of us are very skeptical of string theory and open to
alternatives, since there is absolutely no experimental evidence in
string theory's favor. (To be fair, there is almost no experimental
evidence against it, either. We simply don't know how to probe
things that small.) However, the suggestion that the universe is
a gigantic automaton, with space-time being essentially discrete, 
doesn't hold up, either. 

If the universe were built on a lattice, then the directions of that lattice
could be detected from physical phenomena occurring near the speed of light,
and in particular by the propagation of light itself. If the universe were
built with random local geometry, then light would not have a precise speed, 
and different parts of a signal would travel at slightly different speeds and
directions, much as a prism splits light into differently colored beams. If 
the universe were modeled on an aperiodic tiling with rotational symmetry in a 
statistical sense, there would still be preferred directions at the scale 
of actual experiments. 

All of the simple explanations have failed us. Platonic space-time works
very well for day-to-day life, but the details of the actual universe are
more complicated and mysterious that our human intelligences can
currently fathom.  Not because humans are stupid, but because we have the 
privilege of living in a 
universe of awe-inspiring subtlety and splendor.

Enjoy the ride.

\end{document}